\documentclass[showpacs,preprintnumbers,aps]{revtex4}
\begin{document}
\preprint{USM-TH-165}
\title{On external backgrounds and linear potential in three dimensions}
\author{Patricio Gaete}
\email {patricio.gaete@usm.cl} \affiliation{Departamento de
F\'{\i}sica, Universidad T\'ecnica F. Santa Mar\'{\i}a,
Valpara\'{\i}so, Chile}
\date{\today}
\begin{abstract}
For a three-dimensional theory with a coupling $\phi \varepsilon
^{\mu \nu \lambda } v_\mu  F_{\nu \lambda }$, where $v_\mu$ is an
external constant background, we compute the interaction potential
within the structure of the gauge-invariant but path-dependent
variables formalism. While in the case of a purely timelike vector
the static potential remains Coulombic, in the case of a purely
spacelike vector the potential energy is the sum of a Bessel and a
linear potentials, leading to the confinement of static charges.
This result may be considered as another realization of the known
Polyakov's result.

\end{abstract}
\maketitle

\section{INTRODUCTION}

It is presently widely accepted that quantitative understanding of
confinement remains as the major challenge in $QCD$. In this
respect, the distinction between the apparently related phenomena
of screening and confinement have been of considerable importance
in order to gain further insight and theoretical guidance into
this problem. It is worth recalling at this stage that field
theories which yield a linear potential are relevant to particle
physics, since those theories may be used to understand the
confinement of quarks and be considered as effective theories of
$QCD$. According to this viewpoint, a simple effective theory in
which confining potentials are obtained when a scalar field $\phi$
is coupled to gauge fields via the $3+1$ dimensional interaction
term
\begin{equation}
{\cal L}_I = \frac{g}{8}\phi \varepsilon ^{\mu \nu \alpha \beta }
F_{\mu \nu } F_{\alpha \beta }, \label{INT10}
\end{equation}
has been investigated in Ref.\cite{GaeteG}. The phenomenology of
this model has shown that in the case of a constant electric field
strength expectation value the static potential remains Coulombic,
while in the case of a constant magnetic field strength the
potential energy is the sum of a Yukawa and a linear potential,
leading to the confinement of static charges. More interestingly,
similar results have been obtained in the context of the dual
Ginzburg-Landau theory \cite{Suganuma}, as well as for a theory of
antisymmetric tensor fields that results from the condensation of
topological defects as a consequence of the Julia-Toulouse
mechanism \cite{GaeteW}. Thus, from a phenomenological point of
view, there is a class of models which are good candidates for
effective theories of $QCD$.

In order to give continuity to this program it would be
interesting to verify the above scenario when a scalar field
$\phi$ is coupled to gauge fields via the $2+1$ dimensional
interaction term
\begin{equation}
{\cal L}_I = \frac{g}{2}\phi \varepsilon ^{\mu \nu \lambda } v_\mu
F_{\nu \lambda }, \label{INT15}
\end{equation}
where $v_\mu$ is an external constant background. Our motivation
is mainly to explore the effects of the external background
constant on the confining and screening nature of the potential
that could be useful for later developments in more realistic
theories, such as describing low-energy properties in condensed
matter physics \cite{Nogueira}. Here we should mention that the
interaction term (\ref{INT15}) arises from the dimensional
reduction of Maxwell electrodynamics with the
($Lorentz$-violating) $Carroll$-$Field$-$Jackiw$ term
\cite{Jackiw,Helayel}. The $Lorentz$ violating theories have been
extensively discussed in the last few years. For example, in
connection with ultra-high energy cosmic rays \cite{KosteleckyM},
with space-time varying coupling constants \cite{KosteleckyL}, and
in supersymmetric $Lorentz$-violating extensions \cite{Belich}. In
this Letter, however, we examine the effects of the $Lorentz$
violating background on the interaction energy along the lines of
Refs.\cite{GaeteG,GaeteW}. It must now be observed that this
approach provides a physically-based alternative to the usual
Wilson loop approach, where in the latter the usual qualitative
picture of confinement in terms of an electric flux tube linking
quarks emerges naturally. As we shall see, the static potential
remains Coulombic for a purely timelike vector $v^\mu$. On the
other hand, adopting a purely spacelike vector $v^\mu$, the
potential energy is the sum of a Bessel and a linear potential,
that is, the confinement between static charges is obtained. In
this last respect we recall that, almost twenty years ago,
$Polyakov$ \cite{Polyakov} showed that compact Maxwell theory in
$2+1$ dimensions confines permanently electric test charges. In
this way our calculation provides a complementary phenomenological
picture of the known Polyakov's result, in the hope that this will
be helpful to understand better effective gauge theories in $2+1$
dimensions.

\section{Interaction Energy}

As mentioned above, our principal purpose is to calculate
explicitly the interaction energy between static pointlike sources
for a model one which contains the term (\ref{INT15}). This model
is similar to the studied in Ref.\cite{GaeteG}. To this end we
will compute the expectation value of the energy operator $H$ in
the physical state $\left| \Phi \right\rangle$ describing the
sources, which we will denote by $ \left\langle H \right\rangle
_\Phi$. The Abelian gauge theory we are considering is defined by
the following generating functional in three-dimensional
spacetime:
\begin{equation}
{\cal Z} = \int {{\cal D}\phi {\cal D}A\exp \left\{ {i\int {d^3
x{\cal L}} } \right\}}, \label{TRI10}
\end{equation}
where the Lagrangian density is given by
\begin{equation}
{\cal L} =  - \frac{1}{4}F_{\mu \nu }^2  + \frac{g}{2}\phi
\varepsilon ^{\mu \nu \lambda } v_\mu  F_{\nu \lambda }  +
\frac{1}{2}\partial _\mu  \phi \partial ^\mu  \phi  - \frac{{m^2
}}{2}\phi ^2, \label{TRI15}
\end{equation}
and $m$ is the mass for the scalar field $\phi$. As in
\cite{GaeteG} we restrict ourselves to static scalar fields, a
consequence of this is that one may replace $\Delta \phi = -
\nabla ^2\phi$, with $\Delta \equiv\partial _\mu
\partial ^\mu$. It also implies that, after performing the integration
over $\phi$ in ${\cal Z}$, the effective Lagrangian density is
given by
\begin{equation}
{\cal L} =  - \frac{1}{4}F_{\mu \nu }^2  - \frac{{g^2
}}{8}\varepsilon ^{\mu \nu \lambda } v_\mu  F_{\nu \lambda }
\frac{1}{{\nabla ^2  - m^2 }}\varepsilon ^{\sigma \gamma \beta }
v_\sigma  F_{\gamma \beta }. \label{TRI20}
\end{equation}

By introducing $V^{\nu \lambda}  \equiv \varepsilon ^{\mu \nu
\lambda } v_\mu $, expression (\ref{TRI20}) then becomes
\begin{equation}
{\cal L} =  - \frac{1}{4}F_{\mu \nu }^2  - \frac{{g^2 }}{8}V^{\nu
\lambda } F_{\nu \lambda } \frac{1}{{\nabla ^2  - m^2 }}V^{\gamma
\beta } F_{\gamma \beta }. \label{TRI25}
\end{equation}
At this stage we note that (\ref{TRI25}) has the same form as the
corresponding effective Lagrangian density in four-dimensional
spacetime. This common feature provides the starting point for the
examination of the effects of the $Lorentz$ violating background
on the interaction energy.

\subsection{Spacelike background case}

We now proceed to obtain the interaction energy in the $V^{0i} \ne
0$ and $V^{ij}=0$ ($v_0=0$) case (referred to as the spacelike
background in what follows), by computing the expectation value of
the Hamiltonian in the physical state $\left| \Phi \right\rangle$.
Lagrangian density (\ref{TRI25}) then becomes
\begin{equation}
{\cal L} =  - \frac{1}{4}F_{\mu \nu }^2  - \frac{{g^2 }}{2}V^{0i}
F_{0i} \frac{1}{{\nabla ^2  - m^2 }}V^{0k} F_{0k}  - A_0 J^0,
\label{TRI30}
\end{equation}
where $J^0$ is the external current, $(\mu ,\nu  = 0,1,2)$ and
$(i,k= 1,2)$.

Once this is done, the canonical quantization in the manner of
$Dirac$ yields the following results. The canonical momenta are
\begin{equation}
\Pi^0=0, \label{TRI35}
\end{equation}
and
\begin{equation}
\Pi _i  = D_{ij} E_j, \label{TRI40}
\end{equation}
where $E_i  \equiv F_{i0}$ and $D_{ij}  \equiv \left( {\delta
_{ij} - g^2 V_{i0} \frac{1}{{ \nabla ^2  - m^2 }}V_{j0} }
\right)$. Since $D$ is a nonsingular matrix $(\det D = 1 - g^2
\frac{{{\bf V}^2 }}{{ \nabla ^2  - m^2 }} \ne 0)$ with ${\bf V}^2
\equiv V^{i0} V^{i0}$, there exists the inverse of $D$ and from
Eq.(\ref{TRI40}) we obtain
\begin{equation}
E_i  = \frac{1}{{\det D}}\left\{ {\delta _{ij} \det D + g^2 V_{i0}
\frac{1}{{ \nabla ^2  - m^2 }}V_{j0} } \right\}\Pi _j.
\label{TRI45}
\end{equation}

The corresponding canonical Hamiltonian is thus
\begin{equation}
H_C  = \int {d^2 } x\left\{ { - A_0 \left( {\partial _i \Pi ^i  -
J^0 } \right) + \frac{1}{2}{\bf \Pi} ^2  + \frac{{g^2
}}{{2}}\frac{{\left( {{\bf V} \cdot {\bf \Pi} } \right)^2
}}{{\left( { \nabla ^2  - M^2 } \right)}} + \frac{1}{2} B^2 }
\right\}, \label{TRI50}
\end{equation}
where $M^2\equiv m^2  + g^2 V^2$ and $B$ is the magnetic field.
Requiring the primary constraint (\ref{TRI35}) to be stationary,
leads to the secondary constraint $\Gamma _1 \left( x \right)
\equiv\partial _i \Pi ^i - J^0 = 0$. It is easily verified that
the preservation of $\Gamma_1$ for all times does not give rise to
any further constraints. The theory is thus seen to possess only
two constraints, which are first class. The extended Hamiltonian
that generates translations in time then reads $H = H_C  + \int
{d^2 } x\left( {c_0 \left( x \right)\Pi _0 \left( x \right) + c_1
\left( x \right)\Gamma _1 \left( x \right)} \right)$, where $c_0
\left( x \right)$ and $c_1 \left( x \right)$ are the Lagrange
multiplier fields. Since $ \Pi_0 = 0$ for all time and $ \dot{A}_0
\left( x \right) = \left[ {A_0 \left( x \right),H} \right] = c_0
\left( x \right)$, which is completely arbitrary, we discard $ A_0
\left( x \right)$ and $ \Pi _0 \left( x \right)$ because they add
nothing to the description of the system. Then, the Hamiltonian
takes the form
\begin{equation}
H = \int {d^2 x} \left\{ {\frac{1}{2}{\bf \Pi} ^2  + \frac{{g^2
}}{{2}}\frac{{\left( {{\bf V} \cdot {\bf \Pi} } \right)^2 }}{{
(\nabla ^2 - M^2) }} + \frac{1}{2} B^2  + c(x)\left( {\partial _i
\Pi ^i - J^0 } \right)} \right\}, \label{TRI55}
\end{equation}
where $c(x) = c_1 (x) - A_0 (x)$.

The quantization of the theory requires the removal of
non-physical variables, which is done by imposing a gauge
condition such that the full set of constraints becomes second
class. A convenient choice is found to be \cite{Pato}
\begin{equation}
\Gamma _2 \left( x \right) \equiv \int\limits_{C_{\xi x} } {dz^\nu
} A_\nu \left( z \right) \equiv \int\limits_0^1 {d\lambda x^i }
A_i \left( {\lambda x} \right) = 0, \label{TRI60}
\end{equation}
where  $\lambda$ $(0\leq \lambda\leq1)$ is the parameter
describing the spacelike straight path $ x^i = \xi ^i  + \lambda
\left( {x - \xi } \right)^i $, and $ \xi $ is a fixed point
(reference point). There is no essential loss of generality if we
restrict our considerations to $ \xi ^i=0 $. The Dirac brackets
can now be obtained and the nontrivial Dirac bracket involving the
field variables takes the form
\begin{equation}
\left\{ {A_i \left( x \right),\Pi ^j \left( y \right)} \right\}^ *
=\delta{ _i^j} \delta ^{\left( 2 \right)} \left( {x - y} \right) -
\partial _i^x \int\limits_0^1 {d\lambda x^j } \delta ^{\left( 2
\right)} \left( {\lambda x - y} \right). \label{TRI65}
\end{equation}

We are now in a position to evaluate the interaction energy
between pointlike sources in the model under consideration, where
a fermion is localized at ${\bf y}\prime$ and an antifermion at $
{\bf y}$. From our above discussion, we see that $\left\langle H
\right\rangle _\Phi$ reads
\begin{equation}
\left\langle H \right\rangle _\Phi   = \left\langle \Phi
\right|\int {d^2 x} \left\{ {\frac{1}{2}{\bf \Pi} ^2  + \frac{{g^2
}}{{2}}\frac{{\left( {{\bf V} \cdot {\bf \Pi} } \right)^2 }}{{
(\nabla ^2 - M^2) }} + \frac{1}{2} B^2 } \right\}\left| \Phi
\right\rangle. \label{TRI70}
\end{equation}
Next, as was first established by Dirac\cite{Dirac2}, the physical
state can be written as
\begin{equation}
\left| \Phi  \right\rangle  \equiv \left| {\overline \Psi  \left(
\bf y \right)\Psi \left( {\bf y}\prime \right)} \right\rangle  =
\overline \psi \left( \bf y \right)\exp \left(
{ie\int\limits_{{\bf y}\prime}^{\bf y} {dz^i } A_i \left( z
\right)} \right)\psi \left({\bf y}\prime \right)\left| 0
\right\rangle, \label{TRI75}
\end{equation}
where $\left| 0 \right\rangle$ is the physical vacuum state and
the line integral appearing in the above expression is along a
spacelike path starting at ${\bf y}\prime$ and ending at $\bf y$,
on a fixed time slice. From this we see that the fermion fields
are now dressed by a cloud of gauge fields. From the foregoing
Hamiltonian discussion, we first note that
\begin{equation}
\Pi _i \left( x \right)\left| {\overline \Psi  \left( \bf y
\right)\Psi \left( {{\bf y}^ \prime  } \right)} \right\rangle  =
\overline \Psi  \left( \bf y \right)\Psi \left( {{\bf y}^ \prime }
\right)\Pi _i \left( x \right)\left| 0 \right\rangle  + e\int_
{\bf y}^{{\bf y}^ \prime  } {dz_i \delta ^{\left( 3 \right)}
\left( {\bf z - \bf x} \right)} \left| \Phi \right\rangle.
\label{TRI80}
\end{equation}
Combining Eqs.(\ref{TRI70}) and (\ref{TRI80}), we have
\begin{equation}
\left\langle H \right\rangle _\Phi   = \left\langle H
\right\rangle _0  + V^{\left( 1 \right)}  + V^{\left( 2 \right)},
\label{TRI85}
\end{equation}
where $\left\langle H \right\rangle _0  = \left\langle 0
\right|H\left| 0 \right\rangle$, and the $V^{\left( 1 \right)}$
and $V^{\left( 2 \right)}$ terms are given by:
\begin{equation}
V^{\left( 1 \right)}  =  - \frac{{e^2 }}{2}\int {d^2 x} \int_{\bf
y}^{\bf y^\prime} {dz^\prime_i } \delta ^{\left( 2 \right)} \left(
{x - z^\prime} \right)\frac{1}{{\nabla _x^2  - M^2 }}\nabla _x^2
\int_{\bf y}^{\bf y^\prime} {dz^i } \delta ^{\left( 2 \right)}
\left( {x - z} \right), \label{TRI90}
\end{equation}
and
\begin{equation}
V^{\left( 2 \right)}  =   \frac{{e^2 m^2}}{2}\int {d^2 x}
\int_{\bf y}^{\bf y^\prime} {dz^\prime_i } \delta ^{\left( 2
\right)} \left( {x - z^\prime} \right)\frac{1}{{\nabla _x^2  - M^2
}} \int_{\bf y}^{\bf y^\prime} {dz^i } \delta ^{\left( 2 \right)}
\left( {x - z} \right), \label{TRI95}
\end{equation}
where the integrals over $z^i$ and $z^\prime_i$ are zero except on
the contour of integration.

The $V^{\left( 1 \right)}$ term may look peculiar, but it is just
the familiar Bessel interaction plus self-energy terms. In effect,
expression (\ref{TRI90}) can also be written as
\begin{equation}
V^{\left( 1 \right)}  = \frac{{e^2 }}{2}\int_{\bf y}^{{\bf
y}^{\prime}  } {dz_i^{\prime}}\partial _i^{z^{\prime}} \int_{\bf
y}^{{\bf y}^{\prime}} {dz^i }\partial _z^i G\left( {{\bf
z}^{\prime},{\bf z}} \right), \label{TRI100}
\end{equation}
where $G$ is the Green function
\begin{equation}
G({\bf z}^{\prime}  ,{\bf z}) = \frac{1}{{2\pi }}K_0 \left(
{M|{\bf z}^{\prime}  - {\bf z}  |} \right). \label{TRI105}
\end{equation}
Employing Eq. (\ref{TRI105}) and remembering that the integrals
over $z^i$ and $z_i^{\prime}$ are zero except on the contour of
integration, expression (\ref{TRI100}) reduces to the familiar
Bessel interaction after subtracting the self-energy terms, that
is,
\begin{equation}
V^{\left( 1 \right)}  = - \frac{e^2}{{2\pi }}K_0 \left( {M|{\bf y}
- {\bf y}^{\prime}  |} \right). \label{TRI110}
\end{equation}

The task is now to evaluate the $V^{\left( 2 \right)}$ term, which
is given by
\begin{equation}
V^{\left( 2 \right)}  = \frac{{e^2 m^2 }}{2}\int_{\bf y}^{{\bf
y}^{\prime}  } {dz^{{\prime} i} } \int_{\bf y}^{{\bf y}^{\prime} }
{dz^i } G({\bf z}^{\prime} ,{\bf z}). \label{TRI115}
\end{equation}
By using the integral representation of the Green function
(\ref{TRI105})
\begin{equation}
K_0 \left( x \right) = \int\limits_0^\infty  {\cos (x\sinh t)dt =
} \int\limits_0^\infty  {\frac{{\cos (xt)}}{{\sqrt {t^2  + 1} }}}
dt, \label{TRI120}
\end{equation}
where $x>0$, expression (\ref{TRI115}) can also be written as
\begin{equation}
V^{\left( 2 \right)}   = \frac{{e^2 m^2 }}{{2\pi M^2
}}\int\limits_0^\infty  {dt\frac{1}{{t^2 }}} \frac{1}{{\sqrt {t^2
+ 1} }}\left( {1 - \cos \left( {MLt} \right)} \right),
\label{TRI125}
\end{equation}
where $L\equiv|{\bf y}-{\bf {y^\prime}}|$.

Now let us calculate integral (\ref{TRI125}). For this purpose we
introduce a new auxiliary parameter $\varepsilon$ by making in the
denominator of integral (\ref{TRI125}) the substitution
$t^2\rightarrow t^2+\varepsilon^2$. Thus it follows that
\begin{equation}
V^{\left( 2 \right)} \equiv \lim _ {\varepsilon  \to 0}
{\widetilde V}^{\left( 2 \right)}= \lim _{\varepsilon \to
0}\frac{{e^2 m^2 }}{{2\pi M^2 }}\int\limits_0^\infty
{\frac{{dt}}{{t^2  + \varepsilon ^2 }}} \frac{1}{{\sqrt {t^2  + 1}
}}\left( {1 - \cos \left( {MLt} \right)} \right). \label{TRI130}
\end{equation}
A direct computation on the $t$-complex plane yields
\begin{equation}
{\widetilde V}^{\left( 2 \right)} = \frac{{e^2 m^2 }}{{4 M^2 }}
\left( {\frac{{1 - {\mathop{\rm e}\nolimits} ^{ - ML\varepsilon }
}}{\varepsilon }} \right)\frac{1}{{\sqrt {1 - \varepsilon ^2 } }}
. \label{TRI135}
\end{equation}
Taking the limit $\varepsilon  \to 0$, expression (\ref{TRI135})
then becomes
\begin{equation}
V^{\left( 2 \right)}  = \frac{{e^2 m^2 }}{{4M }}|{\bf y} - {{\bf
y}^\prime}|. \label{TRI140}
\end{equation}
From Eqs.(\ref{TRI110}) and (\ref{TRI140}), the corresponding
static potential for two opposite charges located at ${\bf y}$ and
${\bf y^\prime}$ may be written as
\begin{equation}
V(L) = - \frac{{e^2 }}{{2\pi }}K_0 \left( {ML} \right) +
\frac{{e^2 m^2 }}{{4M}}L, \label{TRI145}
\end{equation}
where $L\equiv|{\bf y}-{\bf {y^\prime}}|$.

It must now be observed that the rotational symmetry is restored
in the resulting form of the potential, although the external
background breaks the isotropy of the problem in a manifest way.
It should be remarked that this feature is also shared by the
corresponding four-dimensional spacetime interaction energy.

We further note that the result (\ref{TRI145}) agrees with that of
Polyakov based on the monopole plasma mechanism, in the short
distance regime. In this way the above analysis reveals that,
although both models are different, the physical content is
identical in the short distance regime. This behavior is also
obtained in the context of the condensation of topological defects
\cite{GaeteW2}.

\subsection{Timelike background case}

Now we focus on the case $V^{0i}=0$ and $V^{ij}\ne 0$ ($v_0\neq0$)
(referred to as the timelike background in what follows). The
corresponding Lagrangian density reads
\begin{equation}
{\cal L} =  - \frac{1}{4}F_{\mu \nu } F^{\mu \nu }  - \frac{{g^2
}}{{8}}V^{ij} F_{ij} \frac{1}{{ \nabla ^2  - m^2 }}V^{kl} F_{kl} -
A_0 J^0, \label{TRI150}
\end{equation}
$(\mu ,\nu  = 0,1,2)$ and $(i,j,k,l = 1,2)$.

Here again, the quantization is carried out using Dirac's
procedure. We can thus write the canonical momenta
$\Pi^\mu=F^{\mu0}$, which results in the usual primary constraint
$\Pi^0=0$ and $\Pi^i=F^{i0}$. Defining the electric and magnetic
fields, as usual, by $ E^i = F^{i0}$ and $B^2  = \frac{1}{2}
F_{ij}F^{ij}$, the canonical Hamiltonian is thus
\begin{equation}
H_C  = \int {d^2 } x\left\{ {\frac{1}{2}{\bf E}^2+\frac{1}{2} B^2
- \frac{{g^2}}{{16}}\varepsilon_{ijm}\varepsilon_{kln}V^{ij} B^{m}
\frac{1}{{ \nabla ^2  - m^2 }}V^{kl} B^{n}  - A_0 \left( {\partial
_i \Pi ^i - J^0 } \right)} \right\}.  \label{TRI155}
\end{equation}
Time conservation of the primary constraint $\Pi^0=0$ leads to the
secondary constraint $\Gamma_1(x) \equiv \partial_i\Pi^i - J^0=0$,
and the preservation of  $\Gamma_1(x)$ for all times does not give
rise to any further constraints. Moreover, it is straightforward
to see that the constrained structure for the gauge field is
identical to the usual Maxwell theory. However, in order to put
the discussion into the context of this paper, it is convenient to
summarize the relevant aspects of the analysis described
previously\cite{Pat}. Therefore, we pass now to the calculation of
the interaction energy.

As in the previous subsection, our objective will be to calculate
the expectation value of the Hamiltonian in the physical state
$\left| \Phi \right\rangle$ (Eq. (\ref{TRI75})). In other words,
\begin{equation}
 \left\langle H \right\rangle _\Phi   = \left\langle \Phi
\right|\int {d^2 x} \left\{ {\frac{1}{2}{\bf E}^2 } \right\}\left|
\Phi \right\rangle. \label{TRI160}
\end{equation}
Then using the above Hamiltonian structure, we obtain the
following form:
\begin{equation}
\left\langle H \right\rangle _\Phi   = \left\langle H
\right\rangle _0  +  \ \frac{{e^2 }}{2}\int {d^2 x\left(
{\int\limits_{\bf y }^{\bf y^\prime} {dz_i \delta ^{(2)} \left( {x
- z} \right)} } \right)} ^2 , \label{TRI165}
\end{equation}
where $\langle H\rangle _{0}=\langle 0\mid H\mid 0\rangle$ and, as
before, the integrals over $z_{i}$ are zero except on the contour
of integrations. We also draw attention to the fact that the
second term on the right-hand side of Eq.(\ref{TRI165}) is clearly
dependent on the distance between the external static fields. In
fact, this term can be manipulated in a similar manner to that in
the three-dimensional case \cite{Pato2}. Accordingly, the
potential for two opposite charges located at $\bf y$ and $\bf y
\prime$ reads
\begin{equation}
V = \frac{{e^2 }}{{2\pi }}\ln(\mu L), \label{TRI170}
\end{equation}
where $\mu$ is a massive cutoff introduced to regularize the
potential, and $L\equiv|\bf y - \bf y^ \prime|$.

\section{Final Remarks}

In summary, we have considered the confinement versus screening
issue for a three-dimensional theory with a coupling $\phi
\varepsilon ^{\mu \nu \lambda } v_\mu  F_{\nu \lambda }$, when the
external constant vector $v_\mu$ is pure spacelike or timelike,
respectively.

It was shown that in the case when the vector $v_\mu$ is purely
timelike no unexpected features are found. Indeed, the resulting
static potential remains Coulombic. More interestingly, it was
shown that when the vector $v_\mu$ is purely spacelike the static
potential displays a Bessel piece plus a linear confining piece.
Effectively, therefore, the model studied leads to a confining
potential between static charges for spacelike $v_\mu$. An
analogous situation in the four-dimensional spacetime case may be
recalled \cite{GaeteG}, where a constant expectation value for the
gauge field strength $\left\langle {F_{\mu \nu } } \right\rangle$
characterizes the external background constant $v_\mu$. Also, a
common feature of these models (three and four-dimensional) is
that the rotational symmetry is restored in the resulting
interaction energy.

We conclude by noting that our result can be considered as another
physical realization of the Polyakov's model. However, although
both models lead to confinement, the above analysis reveals that
the mechanism of obtaining a linear potential is quite different.

\section{ACKNOWLEDGMENTS}
I would like to thank G. Cvetic for helpful comments on the
manuscript. Work supported in part by Fondecyt (Chile) grant
1050546.

\end{document}